\documentclass[aip,jcp,amsmath,amssymb,reprint]{revtex4-1}
\pdfoutput=1
\usepackage{graphicx}
\usepackage{subcaption}
\begin{document}
\title{Charge compensation at the interface between the polar NaCl(111) surface and a NaCl aqueous solution}
\author{Thomas Sayer}
\email{tes36@cam.ac.uk}
\author{Chao Zhang} 
\author{Michiel Sprik} 
\affiliation{Department of Chemistry, University of Cambridge, Cambridge CB2 1EW, United Kingdom}

\begin{abstract}
Periodic supercell models of electric double layers formed at the interface between a charged surface and an electrolyte are subject to serious finite size errors and require certain adjustments in the treatment of the long-range electrostatic interactions. In a previous publication (C.~Zhang, M. Sprik, Phys. Rev. B 94, 245309 (2016)) we have shown how this can be achieved using finite field methods. The test system was the familiar simple point charge model of a NaCl aqueous solution confined between two oppositely charged walls. Here this method is extended to the interface between the (111) polar surface of a NaCl crystal and a high concentration NaCl aqueous solution. The crystal is kept completely rigid and the compensating charge screening the polarization can only be provided by the electrolyte. We verify that the excess electrolyte ionic charge at the interface conforms to the Tasker 1/2 rule for compensating charge in the theory of polar rocksalt (111) surfaces.  The interface can be viewed as an electric double layer with a net charge. We define a generalized Helmholtz capacitance $C_\text{H}$ which can be computed by varying the applied electric field.  We find $C_\text{H} = 8.23 \, \mu \mathrm{Fcm}^{-2}$, which should be compared to the  $4.23 \, \mu \mathrm{Fcm}^{-2}$ for the (100) non-polar surface of the same NaCl crystal. This is rationalized by the observation that compensating ions shed their first solvation shell adsorbing as contact ions pairs on the polar surface.
\end{abstract}

\date{\today}
\maketitle

\section{Introduction} \label{sec:intro}

Crystals exposing a face bearing a net charge are intrinsically unstable if, in addition, the unit cell also has a net dipole moment perpendicular to the surface. Such a termination is referred to as a type III polar surface in the classification by Tasker\cite{Tasker1979}. Tasker explained the instability of type III surfaces by showing that the energy diverges with increasing thickness of the crystal (polar catastrophe). Yet, surfaces with type III orientations do occur in nature. The electrostatic instability is avoided by the accumulation of compensating charge which cancels the dipole moment.\cite{Noguera2000} 

Various compensating mechanisms have been observed or suggested. The dominant mechanisms, as reviewed by Noguera,\cite{Noguera2000} are a change of surface composition (non-stoichiometric reconstruction), adsorption of charged species, and electronic reconstruction (charging of gap or defect states). The review of Ref.~\citenum{Noguera2000} is restricted to oxide materials. It was updated in 2008 in collaboration with Goniakowski\cite{Goniakowski2008} and again in 2013, now also including nano objects, such as thin films.\cite{Noguera2013} Thin films are of special interest because some structures can exhibit an unreconstructed polar surface. Because of their small size they can sustain the polarization field driving the instability in larger systems.

Polar surfaces are clearly a challenge for computational methods.  The problem of how to calculate the total energy of model systems was first addressed as a theoretical exercise in the summation of electrostatic interactions arising from an array of point charges\cite{Tasker1979,Nosker1970,Wolf1992} which was followed later by studies involving electronic structure calculations. The favourite model systems are: MgO (111)(rocksalt), free standing\cite{Refson1995,Goniakowski2002,Wander2003,Finocchi2004,Goniakowski2007,Spagnoli2011,Gaddy2014} or deposited on a metal substrate;\cite{Goniakowski2010}, NiO(111)(rocksalt),\cite{Wander2003} and ZnO, either the (111) zinc blende\cite{Goniakowski2007} or (0001) (wurtzite) \cite{Meyer2003} surface.  More complex surfaces that have been modelled include tetragonal ZrO$_2$(110)\cite{Anez2007,Anez2009} and $\alpha$-$\mathrm{Al}_2\mathrm{O}_3 (0001)$\cite{Ruberto2003}. More recently ternary oxides have attracted attention, in particular polar terminations of LaTiO$_3$ and interfaces with SrTiO$_3$.\cite{Bristowe2011,ChenH2015,Vandewalle2015}

Atomistic model systems necessarily have a slab geometry of limited width.  Electric fields in the solid are permitted.\cite{Bengtsson1999} Similar to the thin films of experiment, the finite slab width can therefore mask a polar catastrophe.\cite{Noguera2013} In this contribution we will outline a finite field approach adjusting the electrostatics. The method is inspired by our work on the atomistic modelling of an electric double layer (EDL) formed by a solid in contact with an aqueous solution (electrolyte).\cite{Zhang2016b}  Finite electric fields penetrating the solid are again a serious concern if the solid is an insulating mineral. The surface charge is of chemical origin: the result of exchange (adsorption or desorption) with an ionic solution (electrolyte). The surface charge is compensated by a zone of excess ionic charge on the electrolyte side of the interface. In a macroscopic, semi-infinite solid the net charge in the EDL is zero. However this basic rule of EDL theory can be violated if the system is represented by a slab with surfaces of opposite charge by creating an internal electric field. The residual electric field manifests as a finite, net charge in the EDL. The surface charge is not properly screened by the mobile ions of the solution, instead the slab acts as a nano-capacitor. 

The nanocapacitor effect, while of interest by itself, must be regarded as a finite system size error if the aim is to simulate an interface between a semi-infinite insulator and an electrolyte. The magnitude of the error was investigated in Ref.~\citenum{Zhang2016b} for the familiar  simple point charge (SPC) model of an aqueous NaCl solution confined between two oppositely charged walls. The walls separating the solution from vacuum were only an atomic diameter thick. Since full 3D periodic boundary conditions (PBC) were applied, moving the supercell over half its length generates a different perspective of the same system. Now the vacuum is in the middle and the system can be viewed as parallel plate capacitor. Even for gaps as large as 20 to 100~\AA, the deficit charge in the EDL turns out to be a significant fraction of the surface charge (up to 20\% at 20~\AA).

The results of Ref.~\citenum{Zhang2016b} were interpreted in terms of a simple  continuum (Stern) model of the two EDL's. Within this model the missing EDL charge was found to be linearly correlated with the electric field in the insulator (the vacuum space).  This observation suggested that charge neutrality of the EDL can be restored by applying an external bias field of the correct magnitude to cancel the field in the insulator. We verified that this is indeed the case, applying a classical molecular dynamics (MD) version of the finite field methods developed by Stengel et al.\cite{Stengel2009,Zhang2016a,Zhang2016c} A further prediction of the continuum model was that the compensating field, termed the field of zero net charge (ZNC), is proportional to the inverse capacitance of the EDL. This relation was also confirmed by the atomistic simulation and used to compute the capacitance.\cite{Zhang2016b}

In the present contribution the finite field methodology of Ref.~\citenum{Zhang2016b} is extended to polar interfaces. The idea is again that application of an appropriate external electric field should cancel the internal field associated with the polarization. In this first application the configuration of the ions in the solid is strictly fixed. Only the ions and water molecules in the electrolyte move and should therefore screen the polarization. The question is thus whether the excess charge supplied by the electrolyte can play the role of the compensating charge density in the theory of polar surfaces.\cite{Tasker1979,Noguera2000,Goniakowski2008} The Tasker rule makes a very precise statement about the compensating charge which should be relatively easy to verify in a simple classical point charge model as will be used here.

\section{Model and method} \label{method}

\subsection{Point charge model for the NaCl-electrolyte interface} \label{sec:spc}

\begin{figure}
\begin{center}
	\resizebox{.475\textwidth}{!}{\includegraphics{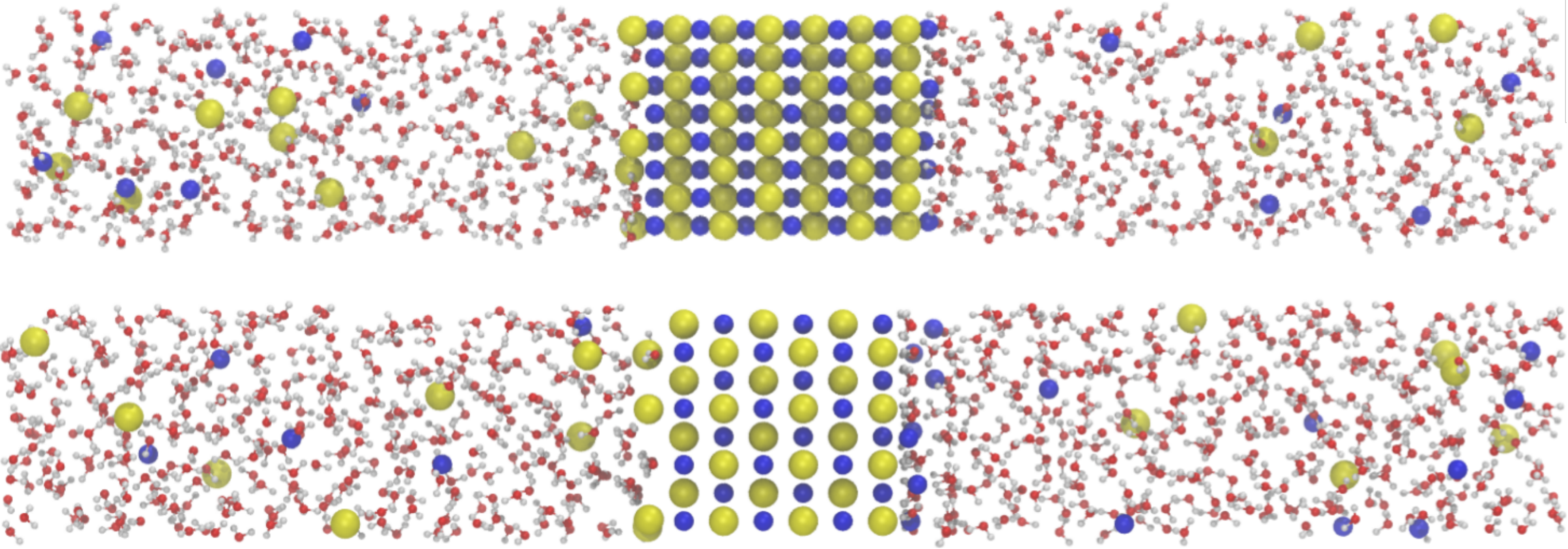}}
		\label{fig:polar}
\end{center}
\caption{\label{fig:renders}
MD snapshots of the polar (111) (top) and non-polar (100)  (bottom) surfaces of a rigid crystalline NaCl slab interfaced with a high concentration NaCl aqueous solution. Na$^+$ ions are depicted in blue, Cl$^-$ ions in yellow. The non-polar (100) surface is given an artificial net charge of  $8e$ matching the surface charge $16e$ of the polar(111) surface. A finite electric field is applied cancelling the internal field of the slab. With the polarization field removed the excess charge in the electrolyte near the surface should compensate the slab polarization.}
\end{figure}

The main objective of this study is validation of the finite field method for electrochemical interfaces of polar surfaces. We opted therefore for the simplest of model systems, a slice of NaCl crystal of (111) orientation. The slab is terminated on one side by a Na$^+$ plane and the other side by a Cl$^-$ plane and is kept rigidly in the bulk rock salt geometry. The solid is in contact with a high concentration aqueous NaCl solution. Fig.~\ref{fig:renders}a shows an instantaneous configuration sampled from the MD simulation. The classical force model is the same as used in Ref.~\citenum{Zhang2016b} (for further detail see section \ref{sec:md}). 

 The theory (Tasker rule) says that the surface charge density required to cancel the dipole generated by cleaving a rocksalt structure along a (111) plane is $-\sigma_0/2$, where $\sigma_0$ is the surface charge density of the terminating plane. A crystal with this solid/electrolyte interface should be stable and have no net internal electric field. The procedure is therefore the same as in Ref.~\citenum{Zhang2016b}. The applied bias field is varied until the internal Maxwell field in the slab is on average zero. We computed the charge imbalance (space charge) in the electrolyte  in contact with the crystal face and checked whether it has the theoretical value of $-\sigma_0/2$. 

\subsection{Stern model for the polar surface-electrolyte interface} \label{sec:stern}

\begin{figure*}
\begin{center}
	\resizebox{0.75\textwidth}{!}{\includegraphics{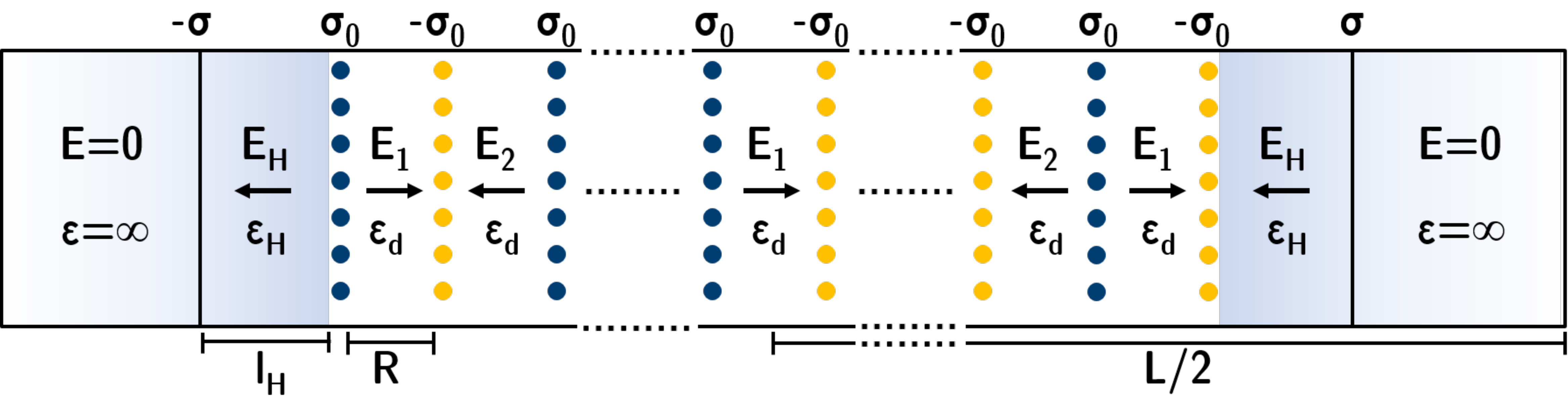}}
\end{center}
\caption{\label{fig:stern}
Schematic drawing of the continuum model of a polar surface-electrolyte system under periodic boundary conditions (PBC). The (absolute) surface charge density of a polar surface is $\sigma_0$ and the compensating charge induced in the electrolyte solution is $\sigma$. The solid slab is separated from the electrolyte on both sides by Helmholtz layers. The dielectric constants of the Helmholtz layers and polar solid are $\epsilon_\text{H}$ and $\epsilon_\text{d}$ respectively. The box size is $L$, the width of Helmholtz layer is $l_\text{H}$ and the thickness of a layer in polar solid is $R$. The arrows indicate the convention for the sign of the uniform electric fields in the Helmholtz layers and crystal segments.}
\end{figure*}

Anticipating our results we found that the Tasker half surface charge density rule for the unreconstructed rocksalt (111) termination is indeed satisfied for our model. This raises questions about the electrostatics of the  polar surface/electrolyte interface. What are the differences compared to the regular charge neutral EDL induced by the surface charge of a non-polar dielectric solid? What is the capacitance or how can we even define a capacitance? We will address this problem by generalizing the continuum Stern model of Ref.~\onlinecite{Zhang2016b}. The model is pictured in Fig.~\ref{fig:stern}.  As before the electrolyte is partitioned in a proper ionic conductor ($\epsilon=\infty$) and two boundary layers of each of thickness $l_{\mathrm{H}}$. The dielectric constant in a boundary layer, to which we will continue to refer as a ``Helmholtz layer'', is $\epsilon_{\mathrm{H}}$. The Maxwell field in the conducting region is strictly zero while it can be finite in a Helmholtz layer. This field is denoted by $E_{\mathrm{H}}$. The polarization of the electrolyte is represented by the surface charge density $\pm \sigma$ of the planes separating the conductor and the Helmholtz layer. With the surface charge of the solid fixed, the electrolyte surface charge $\sigma$ is the central variable in the model.

 To model the polar slab we follow Noguera,\cite{Noguera2000} representing the solid by a succession of planes with alternating surface charge density $\sigma_0$. The planes are a distance $R$ apart. There are $n+1$ of these planes dividing the solid up into $n$ slices, where $n$ is an odd number. The dielectric constant is homogeneous throughout the solid and will again be indicated by $\epsilon_\text{d}$. The electric field is however not the same everywhere. Counting from the left in Fig.~\ref{fig:stern} the field $E_1$ in the first layer ($i=1$) is different from the field $E_2$ in the next layer ($i=2$). Because of the strict alteration of the surface charges on the planes the field in all odd numbered regions is $E_1$ and in the even numbered layers $E_2$ (note the convention of the field directions in Fig.~\ref{fig:stern}). To represent a MD supercell the Stern model is periodically repeated in the normal direction. The length of the supercell is $L$. 

 In the finite field method of Ref.~\onlinecite{Zhang2016b} the periodic MD system is subject to an electric field $\bar{E}$ using an extended Hamiltonian (see section \ref{sec:md}).  Note that $\bar{E}$ is not an external field $E_0$ but the average of the Maxwell field. The product $V = -\bar{E}L$ can therefore be directly interpreted as the potential difference across the MD cell.  The corresponding cell potential $V$  is the sum of all potential differences over the uniform layers making up the system and we can write   
\begin{equation} \label{eq:barE}
\bar{E}L = -2 E_{\mathrm{H}}l_{\mathrm{H}} + n\bar{E}_{d}R,
\end{equation}
with the field in the electrolyte set to zero (see Fig.~\ref{fig:stern}). $n \bar{E}_d R $ is (minus) the potential across the polar solid with average field $\bar{E}_d$ given by   
\begin{equation} \label{eq:Ed}
\bar{E}_\text{d} = -\frac{n-1}{2n}E_{2}+\frac{n+1}{2n}E_{1}.
\end{equation} 

The Maxwell equations for boundaries give the following dependencies:
\begin{eqnarray} 
\epsilon_{\mathrm{H}}E_{\mathrm{H}} + \epsilon_\text{d}E_{1} & = & 4\pi\sigma_{0}
\label{eq:maxwell1} \\[4pt] 
\epsilon_{\mathrm{H}}E_{\mathrm{H}} & = & 4\pi\sigma
\label{eq:maxwell2} \\[4pt] 
\epsilon_\text{d}E_{2} + \epsilon_\text{d}E_{1} & = & 4\pi\sigma_{0}.
\label{eq:maxwell3}
\end{eqnarray}
Exchanging the fields in the right-hand side of Eq.~\ref{eq:barE} for charge densities using Eqs.~\ref{eq:maxwell1}, \ref{eq:maxwell2} and \ref{eq:maxwell3} we obtain an expression for $\sigma$ in terms of $\sigma_0$ and $\bar{E}$,
\begin{equation} \label{eq:sigma_flat}
 \sigma = \left( \frac{n+1}{2} \frac{\sigma_{0}}{C_\text{d}} - \bar{E}L \right)
 \left( \frac{2}{C_{\mathrm{H}}}+\frac{n}{C_\text{d}} \right)^{-1},
\end{equation}
where $C_\text{d}= \epsilon_\text{d}/(4\pi R)$ and 
$C_{\mathrm{H}}= \epsilon_{\mathrm{H}}/(4\pi l_{\mathrm{H}})$. The parameters  $C_\text{d}$ and  $C_{\mathrm{H}}$ have the familiar form of the capacitance of a parallel plate capacitor and will be interpreted as such.  

The surface charge $\sigma_0$ is fixed. The Maxwell field $\bar{E}$ is the only control parameter in the model. $\bar{E}$ can be varied until the average field $\bar{E}_d$  within the crystal is zero. In this state, referred to as the point of ``Compensating Net Charge'' (CNC), the two interfaces of the slab are decoupled, and so the infamous finite size error is removed.  Setting $\bar{E}_d = 0 $ in Eq.~\ref{eq:barE} and inserting into Eq.~\ref{eq:sigma_flat} using Eq.~\ref{eq:maxwell2}  gives the compensating charge provided by the electrolyte:
\begin{equation} \label{eq:sigma_n}
     \sigma = \frac{n+1}{2n}\sigma_{0}.
\end{equation}
Indeed in the limit $n\rightarrow\infty$, the prefactor of equation \ref{eq:sigma_n} tends towards $\frac{1}{2}$ in agreement with the Tasker rule for a rocksalt(111) polar surface.\cite{Tasker1979,Noguera2000} The compensating charge is determined by the surface charge only and is independent of all other structural parameters.    

The $\bar{E}=\bar{E}_{\mathrm{CNC}}$  field plays the same role as the field of zero net charge $\bar{E}_{\mathrm{ZNC}}$ of Ref.~\onlinecite{Zhang2016b}. At $\bar{E} = \bar{E}_{\mathrm{ZNC}}$ the field in the dielectric slab is zero restoring net charge neutrality to the EDL. Moreover the Stern model led to a simple relation between the capacitance of the EDL and $\bar{E}_{\mathrm{ZNC}}$ (Eq.~37 of Ref.~\onlinecite{Zhang2016b})   
\begin{equation} \label{eq:cHedl}
 C_{\mathrm{H}}=\frac{2\sigma_{0}}{\bar{E}_{\mathrm{ZNC}}L}
\end{equation} 
Pursuing the parallel with the ``regular'' EDL of Ref.~\onlinecite{Zhang2016b} further, we can similarly express the capacitance of the Helmholtz layer in terms of $\bar{E}_{\mathrm{CNC}}$. Combining Eq.~\ref{eq:sigma_flat} and \ref{eq:sigma_n}, equivalent to imposing a potential $-\bar{E}_{\mathrm{CNC}}L$, we find
\begin{equation} \label{eq:cH}
 C_{\mathrm{H}}=\frac{2\sigma_{0}}{\bar{E}_{\mathrm{CNC}}L}\frac{n+1}{2n}.
\end{equation}
Recall that the factor 2 multiplying $\sigma_0$ in Eqs.~\ref{eq:cH} and \ref{eq:cHedl} is there because $-L\bar{E}$ is the potential over a pair of EDL's in series each with a capacitance $C_{\mathrm{H}}$.

The status of $C_{\mathrm{H}}$ as the Helmholtz capacitance of the polar surface/electrolyte interface may look ambiguous because the response charge of the electrolyte is only half the surface charge. It is therefore of interest to compare the capacitance for the polar NaCl $(111)$ surface as defined by Eq.~\ref{eq:cH} to the capacitance obtained by charging a non-polar surface. The obvious candidate is the $(100)$ surface. The surface was charged by slightly enhancing the positively (negatively) charged ions compared to the counter charge. The expression for $C_{\mathrm{H}}$ is given by Eq.~\ref{eq:cH} leaving out the factor $(n+1)/(2n)$ consistent with Eq.~\ref{eq:cHedl}.

\subsection{Hamiltonian and finite electric field molecular dynamics} \label{sec:md}

All simulations  were performed under ambient conditions using a modified version of the GROMACS package.\cite{Hess2008} The water model is SPC/Extended.\cite{Berendsen:1987uu} The SPC model of aqueous Na$^+$ and Cl$^-$ was taken from Ref.~\onlinecite{Joung2008} and has been validated for high salt concentrations.\cite{Zhang:2010zh,Zhang:2012fo} (see also the review by Nezbeda et al.~\cite{Nezbeda2016}). Identical force field parameters were used for the interactions with the ions in the (rigid) NaCl crystal. The supercell cross-sectional area were 2.200 and 2.545 nm$^2$ for the (111) and (100) orientation respectively. The corresponding distance between the charged planes of the (111) slab, referred to as $R$ in Fig.~\ref{fig:stern}, is $R=1.63$ \AA. The MD cell lengths were adjusted to keep variations in solvent properties to a minimum. All supercells contained 20 aqueous NaCl ion pairs in $\sim$ 600 waters, evenly dispersed in the initial configuration. This particular concentration was chosen such that the electrolyte remains a good ionic conductor after the formation of EDLs has reduced the concentration (the total number of ions is fixed). The MD parameters were as follows: the NVT ensemble was employed, and a temperature of 298 K mainted by Nos\'{e}-Hoover thermostat with a coupling constant of 0.4 ps;\cite{Martyna1992} the timestep was 2 fs and the simulations were run for a total of 1 ns; the electrostatics were computed using 4$^{th}$ order Particle Mesh Ewald (PME) summation with a Fourier spacing of 0.6 \AA{}  and a real-space cutoff of 6.5 \AA.\cite{Darden1993} The first 200 ps were discarded as equilibration time. The system was considered equilibrated if the electrostatic potential in the bulk electrolyte was flat. Ref~\onlinecite{Zhang2016b} investigated the equilibration time for $C_{\mathrm{H}}$ and the choice of 200 ps achieves this level of convergence in the present work.

The modification of the GROMACS package concerns the implementation of the constant field Hamiltonian method according Vanderbilt and co-workers.\cite{Stengel2009} Forces and energies are derived from the extended Hamiltonian
\begin{equation}    
\label{fhmd}
H_E\left( v, \bar{E} \right) = H_{\mathrm{PBC}}(v)- \mathit{\Omega} \bar{E} P(v),
\end{equation}
where $H_{\textrm{PBC}}(v)$ is the Hamiltonian as defined by the SPC model. $v= (\mathbf{r}^N ,\mathbf{p}^N)$ stands for the collective  momenta and position coordinates of the $N$ particles in the system. The subscript PBC indicates that the electrostatic energies and forces are computed using standard Ewald summation (``tin foil'' boundary conditions). $\mathit{\Omega}$ is the volume of the MD cell. $P$ is the polarization perpendicular to the crystal slab with $\bar{E}$ the magnitude of the electric field. $P$ is computed from the total dipole moment of the supercell including the contribution from the ions in the solution and the solid. Typical values of $\bar{E}$ applied to our model system are in the order of 1 V/nm.   

Finite electric field Hamiltonians of the form of Eq.~\ref{fhmd} are familiar in classical MD.\cite{Yeh1999} As already pointed out by Yeh and Berkowitz, the electric field in the dipole coupling term $-\mathit{\Omega} \bar{E} P$, when combined with Ewald summation, must be interpreted as the average of the Maxwell field rather than the applied external field.\cite{Yeh1999} In Ref.~\onlinecite{Stengel2009} this feature, specific to standard Ewald summation,  is given a firm thermodynamic foundation (see also Ref.~\onlinecite{Zhang2016a}). A further important point is that including ions in the polarization makes $P$ a multi-valued function depending on the supercell boundary. This is a central theme in the modern theory of polarization which applies to electronic polarization as well as classical point charge systems.\cite{Resta2007} This issue becomes critical for a supercell of the geometry of Fig.~\ref{fig:renders}. The mobile ions can cross the cell boundaries. When this happens the ions must be followed out of the cell in the calculation of the polarization. In classical simulation, this definition of polarization was introduced in the context of the MD study of electrolytic solutions and is generally referred to as itinerant polarization\cite{Caillol1994}. Note that the forces derived from the Hamiltonian Eq.~\ref{fhmd} are not affected by ambiguities in the definition of polarization (see Ref.~\onlinecite{Zhang2016b} for a more detailed discussion).

\section{Results} \label{sec:results}

\subsection{Finding the field of compensating net charge (CNC)} \label{sec:cnc}

\begin{figure}
\begin{center}
	\resizebox{.475\textwidth}{!}{\includegraphics{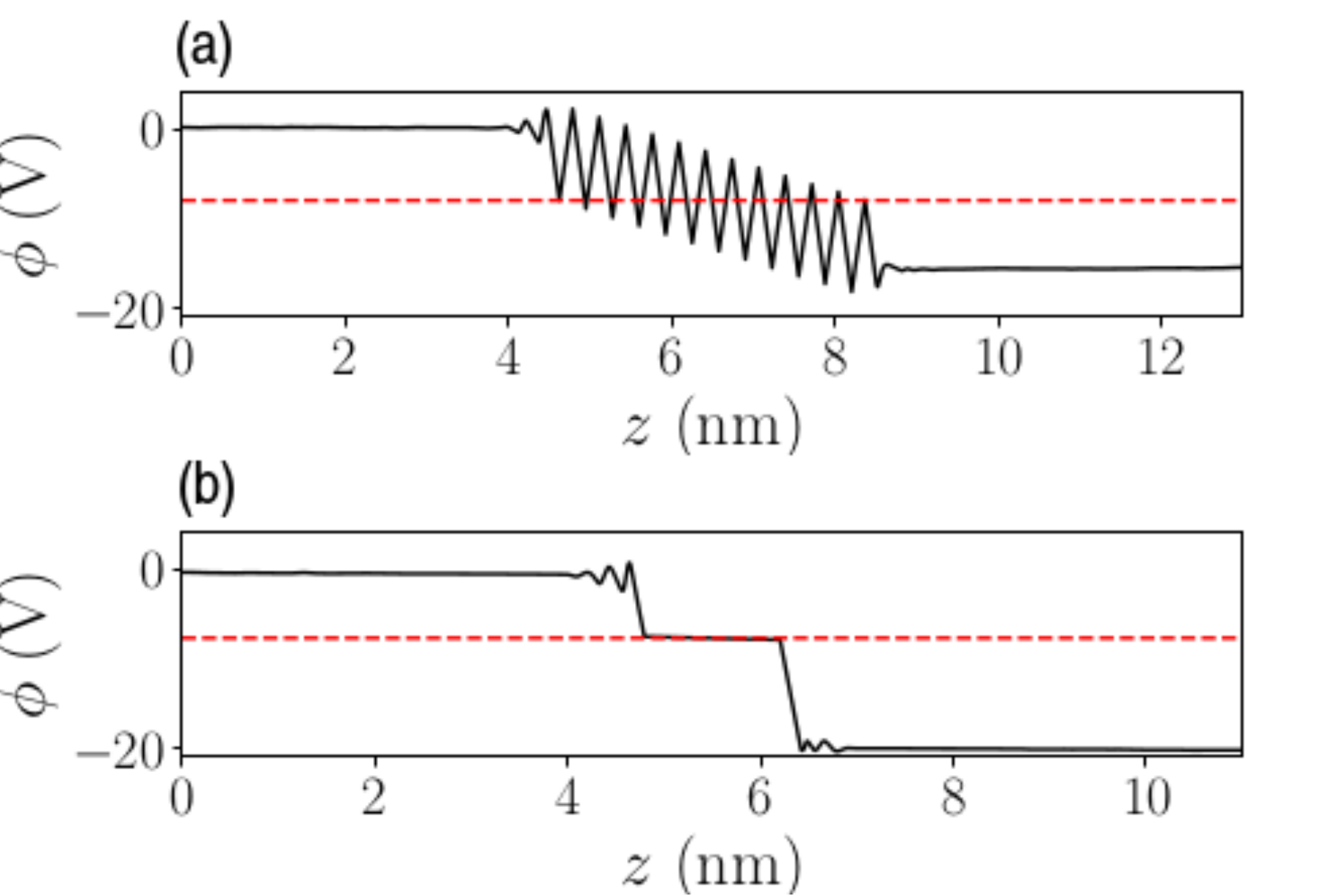}}
		\label{fig:pot1}
\end{center}
\caption{(a) Potential profile (potential vs distance) for the NaCl (111)-electrolyte system for n=23 at $\bar{E}_{\mathrm{CNC}}L =$ 1.2 V. The first and last crystal peak can be seen to occur at the same potential conform the requirement for CNC. (b) Potential profile for the NaCl (100)-electrolyte system with surface charge $8e$ at $\bar{E}_{\mathrm{CNC}}L =$ 1.8 V.}
\label{fig:graphspot}
\end{figure}

The first step is locating the field of compensating net charge. In the state of CNC, $\bar{E}_\text{d}=0$ by definition. The change $\Delta \phi_\text{d}$ in the electrostatic potential over the length of the crystal must therefore be zero. $\Delta \phi_\text{d}$ was computed from the electrostatic potential profile $\phi(z)$. As is evident from Fig.~\ref{fig:graphspot}(a) the potential profile, while flat in the electrolyte,  shows a saw tooth pattern in the crystal reflecting the alternating charge of the (111) crystal planes. In contrast, the potential profile in the non-polar slab (Fig.~\ref{fig:graphspot}(b)) is smooth because (100) planes are net neutral. In a CNC state ($\bar{E}_\text{d}=0$), from which the configuration of Fig.~\ref{fig:graphspot}(b) was sampled, the potential is therefore constant ($\Delta \phi_\text{d} = 0$). 

It is perhaps instructive to analyze the polar profile of Fig.~\ref{fig:graphspot}(a) in somewhat more detail because its appearance may at first seem inconsistent with the condition of vanishing average internal electric field required at CNC. Referring back to Fig.~\ref{fig:stern}  we note that the period of the modulation of the potential in  Fig.~\ref{fig:graphspot}(a) is $2R$. However, the width of the slab is $nR$ where $n$ is an odd integer. The CNC field therefore aligns a maximum of the saw tooth at one face with a minimum at the opposite face as indicated by the red dashed line in the figure.  This can only be achieved by canting the  sequence of maxima (minima) which is the feature that stands out in Fig.~\ref{fig:graphspot}(a). 

$\Delta \phi_\text{d}$ was tracked against the applied field $\bar{E}$ to locate the point of CNC for increasing width of the crystal. The result is shown in Fig.~\ref{fig:montage}(c). The width of the slab is represented by the number $n$  of capacitors in series (see Fig.~\ref{fig:stern}). The figure suggests that a minimum of $n=15$ layers (16 planes) is needed to reach convergence. This corresponds to a slab width of $25$ \AA. In the search for the CNC it was observed that the response of charge and potential to changes in the field was linear. The exception is the terminal n=1 system. This is likely due to dielectric saturation at the high CNC field for n=1 as suggested by Fig.~\ref{fig:montage}(c). Therefore, the values for n=1 are obtained by extrapolation from the low-field, linear regime.  

Slab width is of course always a critical parameter in periodic models of interfaces. Solid NaCl/aqueous NaCl interfaces are popular model systems and the question of size dependence has been investigated in great detail in calculations of solubility from the direct equilibrium between solid and solution\cite{Nezbeda2016}. Using an interaction model identical to the one used here, Espinoza and coworkers found in a careful study\cite{Espinoza2016} that the solubility is essentially converged for slabs of a width larger than 40 \AA. This number exceeds but is still comparable to the $25$ \AA{} inferred from Fig.~\ref{fig:stern}. Considering however, that the orientation of the slab used in Ref.~\citenum{Espinoza2016} is the nonpolar $(100)$ surface, the similarity between the finite size effects is perhaps somewhat surprising. In this context we should also point out that the size effect for our polar interface seems to be captured to great accuracy by an analytical expression (Eq.~\ref{eq:sigma_n}) specified by only a single parameter, the surface charge density $\sigma_0$ of the polar $(111)$ surface for which there is no direct counter part in the dissolution of a $(100)$ interface. On the other hand the consistency in finite size effects can also be interpreted as an indication that surface charge and polarity  play a role in the dissolution of non-polar interfaces.

\subsection{Compensating electrolyte charge and capacitance} \label{sec:edlcap}

\begin{figure}
\centering
\resizebox{0.48\textwidth}{!}{\includegraphics{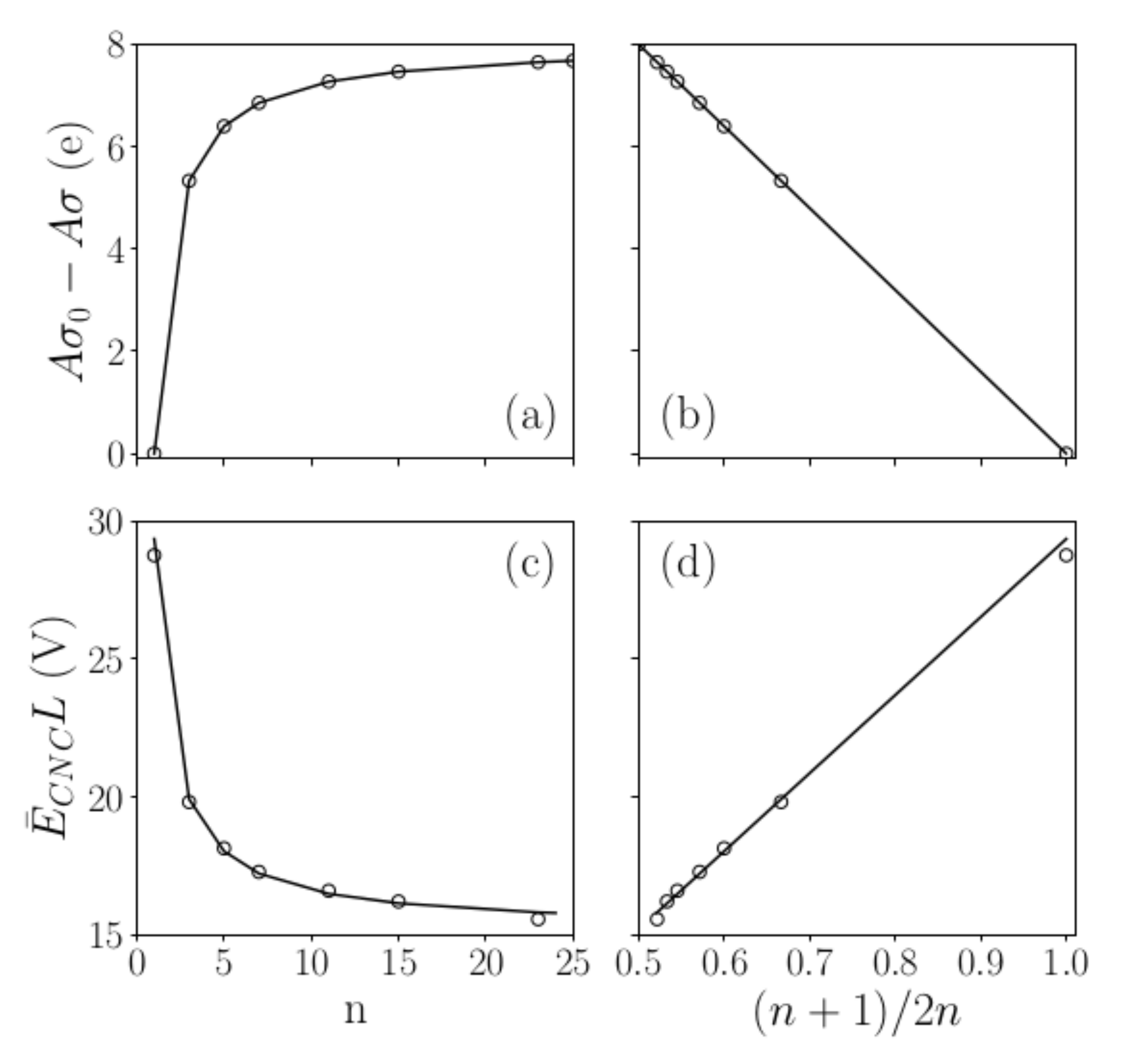}}
\caption{(a) The effective surface charge of the electrolyte at the point of CNC for increasing width $n$ of the slab (see Fig.~\ref{fig:stern}). It is represented as the difference between the total fixed surface charge  $A \sigma_0$ of the crystal and the response charge $A\sigma$ from the electrolyte solution, where $A$ is the area of the polar plane in the supercell. $ A \sigma_0 = 16 e$ in our model.  (b)Plot of $A (\sigma_{0}-\sigma) $ vs $(n+1)/2n$. The slope of $-A \sigma_{0}$ demonstrates that Eq.~\ref{eq:sigma_n} is obeyed. (c) Plot of $\bar{E}_{\mathrm{CNC}}L$ vs $n$. (d) Plot of $\bar{E}_{\mathrm{CNC}}L$ vs $(n+1)/2n$. The slope of $2\sigma_{0}/C_\text{H}$ allows us to extract the value of the Helmholtz capacitance $C_\text{H}$.}
\label{fig:montage}
\end{figure}

In the simple Stern model of Fig.~\ref{fig:stern}, the charge induced in the electrolyte is represented as a surface charge density $\sigma$ at the sharp interface between the electrolyte and the Helmholtz layer. In the atomistic model system of Fig.~\ref{fig:renders}, the interface is more diffuse. We have estimated the compensating charge as the integral of the excess charge density (``space charge'') of the electrolyte in contact with the surface. This procedure is outlined in more detail in Ref.~\onlinecite{Zhang2016b}. The excess charge per unit area is identified with the $\sigma$ of the Stern model.  The results are plotted in Fig.~\ref{fig:montage}(a) where we have represented the estimates of $\sigma$ as total surface charges $ A \sigma$  with $A$ the MD cell cross section. 

The surface charge of a NaCl (111) plane in our model system is $A \sigma_0 = 16e$. The theoretical compensating charge (Tasker rule) is therefore $8e$. Fig.~\ref{fig:montage}(a) shows that the compensating charge $A \sigma$ approaches $8e$ when the number of layers of polar solid gets large enough. Therefore, our simulation confirms that the charge imbalance is in accord with the theoretical value. Indeed, figure \ref{fig:montage}(b) shows excellent agreement with Eq.~\ref{eq:sigma_n} derived from the continuum model Fig.~\ref{fig:stern}.  This also gives us confidence that Eq.~\ref{eq:cH} can be used to extract a value for the Helmholtz capacitance from simulations. Indeed, Fig.~\ref{fig:montage}(d) shows the desired linear (n+1)/2n dependence of Eq.~\ref{eq:cH}. The slopes gives an estimation of $C_\text{H}$ of 8.23 $\mu$Fcm$^{-2}$ for the polar (111) surface.

\begin{figure}
	\centering
	\resizebox{0.4\textwidth}{!}{\includegraphics{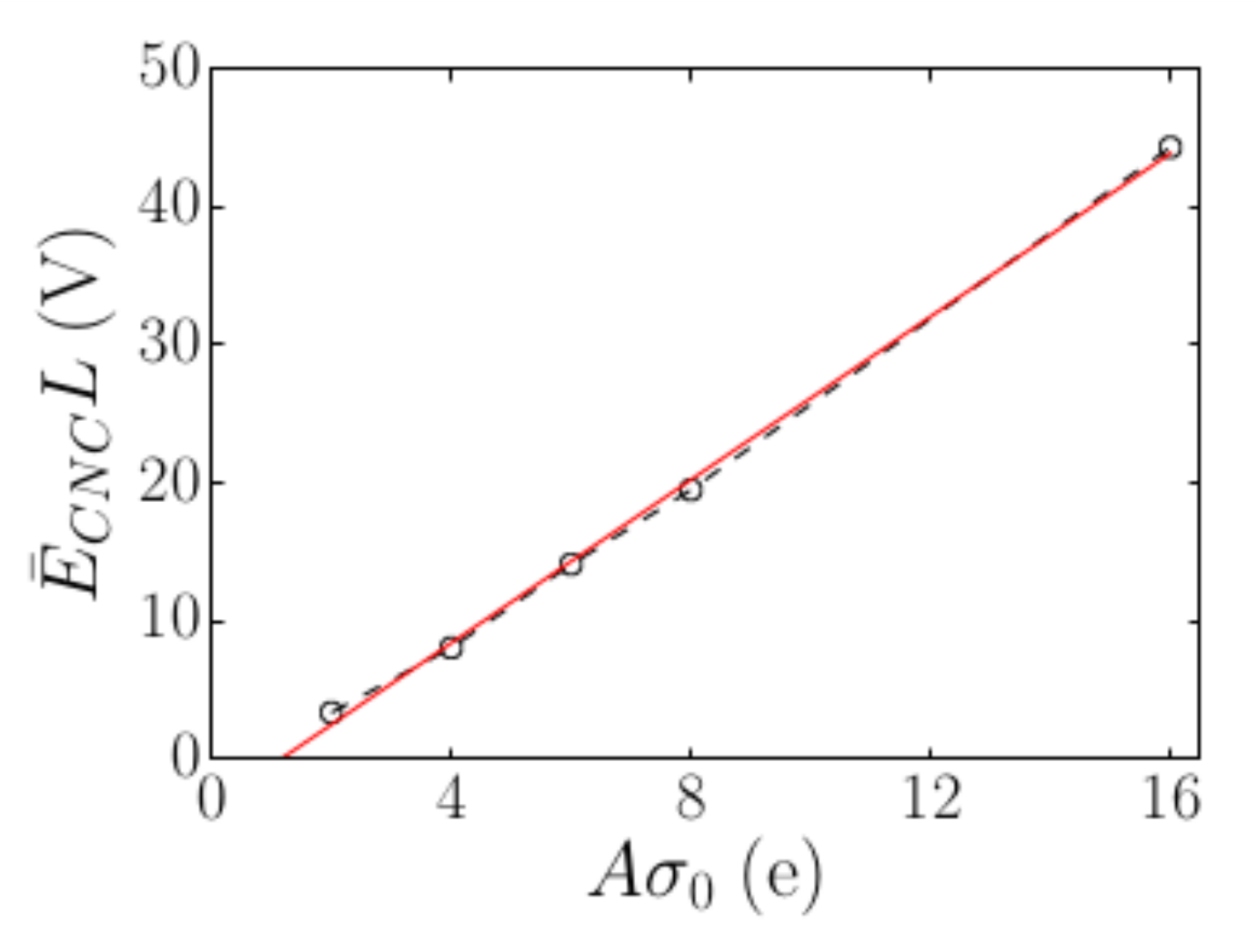}}
        \caption{CNC Potential $\bar{E}_\text{CNC}L$ (is potential of zero net EDL charge)  as a function of the artificially enhanced surface charge for the NaCl (100)-electrolyte system. The slope of the linear fitting (red solid line) is linked to the Helmholtz capacitance $C_\text{H}$ according to Eq.~\ref{eq:cHedl}.}
	\label{fig:chargevznc}
\end{figure}

8.2  $\mu$Fcm$^{-2}$ is a relatively modest capacitance, not much larger than the 4.4  $\mu$Fcm$^{-2}$ we found in Ref.~\onlinecite{Zhang2016b} for the EDL formed  by a 1.4~M  aqueous NaCl electrolyte confined between two walls of uniform opposite surface charge densities.   While our procedure of a fractional increase of the charge of all ions of one species (Na$^+$ or Cl$^-$) is nonphysical, we argue that the electrostatics of an EDL formed by this system is similar to that of the ``regular'' EDL of Ref.~\onlinecite{Zhang2016b}. Accordingly, the Helmholtz capacitance can be obtained from Eq.~\ref{eq:cHedl} or equivalently from Eq.~\ref{eq:cH} by leaving out the $(n+1)/2n$ factor. The EDL of the non-polar surface is charge balanced for vanishing internal field. The CNC and ZNC are interchangeable in this case. The result of the applied voltage at CNC as a function of the surface charge for the non-polar (100) surface is shown in Fig.~\ref{fig:chargevznc}. The slope yields the Helmholtz capacitance $C_\text{H}$ with a value 4.23 $\mu$Fcm$^{-2}$, effectively identical to the capacitance obtained in Ref.~\onlinecite{Zhang2016b}. 

\subsection{Electric double layer structure} \label{sec:edlstruc}

\begin{figure}
	\centering
	\resizebox{0.45\textwidth}{!}
{\includegraphics{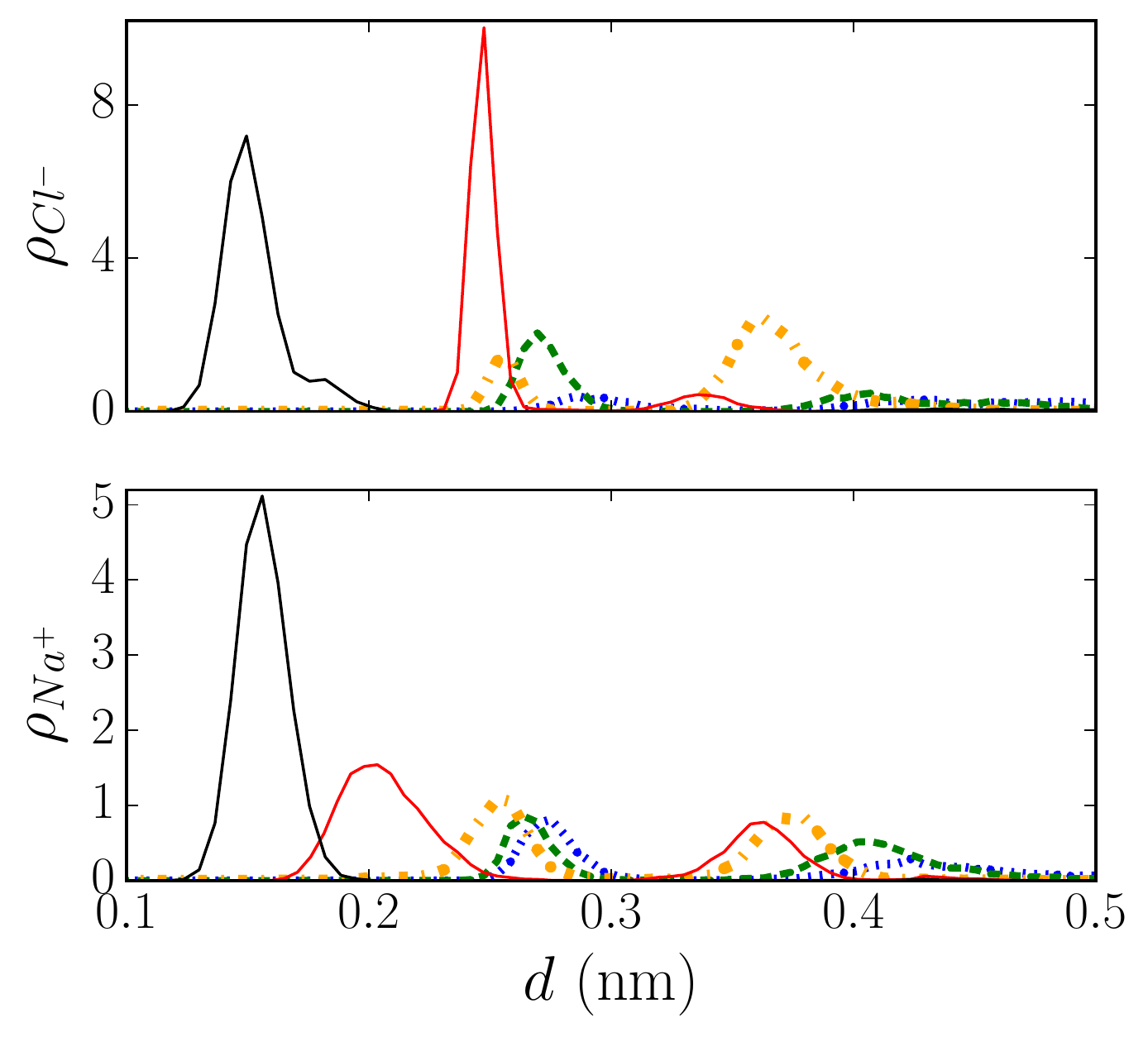}}
       	\caption{Density plot of compensating ions (excess charge) from the electrolytic solution at points of CNC. $d$ is defined as the distance of ions from the relevant surface of NaCl crystals. The solid black line is the density profile for the (111) polar surface with surface charge $16e$. The solid red line is for the (100) non-polar surface with artificial charge $A\sigma_{0}=8e$. The long-dashed orange, short-dashed green, and dotted blue lines are the (100) surface with $A\sigma_{0}$ values of $6e$, $4e$, and $2e$ respectively.}
	\label{fig:densityplot}
\end{figure}

The near 2 to 1 ratio for the capacitance of the polar (111) and non-polar (100) surfaces may not be a coincidence. A possible explanation for this behaviour can be seen in the double layer structure of Fig.~\ref{fig:densityplot}. For the (111) case, the ions of the double layer are adsorbed to the surface at a distance of $\sim$1.5~\AA, shedding their inner solvation shell. The formation of contact ion pairs is particularly noteworthy for the Na$^{+}$ ions, which are traditionally thought to have an almost unbreakable solvation shell. In the (100) case, at the lowest charge of 2e, there is no driving force for such a dehydration, and the first peak in the density occurs at the larger distance of 2.8~\AA{}, followed by a small secondary peak further out. As the surface charge is increased, the profile shifts towards the surface. Even for a (100) surface charge as high as the compensating charge for the (111) polar surface ($8e$), the structural difference between EDLs is apparent. The capacitance of a compact (Helmholtz) EDL is inversely proportional to its width, therefore, the 2 to 1 ratio in $C_{\mathrm{H}}$ between the surfaces is roughly consistent with the relative positions of the counter ions in Fig.~\ref{fig:densityplot}. 

\section{Conclusion and outlook}

Our calculations show that the excess charge in a high concentration NaCl aqueous solution adjacent to a rigid NaCl (111) surface complies to the Tasker rule for a polar surface of this geometry (half the surface charge). This was to be expected. The Tasker rule is based on general considerations involving geometry and electrostatics. It would have been a surprise if this rule would not hold for polar surface electrolyte interfaces. The results of the present study are therefore intended as further validation of our finite field method for the simulation of electric double layers under full periodic boundary conditions. This method was introduced in Ref.~\citenum{Zhang2016b} and applied to a conventional electric double layer for which the electrolyte counter charge equals the opposite of the surface charge (zero net charge). Reproducing the half charge predicted by the Tasker rule for a model polar surface seemed to us a separate challenge which was taken up in the calculation reported here. 

The key function of the applied external field was to compensate for the internal electric field in polar model slabs. Slabs of a width accessible to molecular simulation can sustain an internal field leading to violation of the Tasker rule for polar surfaces of semi-infinite crystals. The fact that the electrolyte is an ionic conductor is essential. This enforces a zero macroscopic field in the bulk region of the electrolyte whatever the magnitude of the polarization of the slab or the applied field. The external field can therefore be adjusted to cancel  the internal field in the solid without inducing a field in the electrolyte.          

The model system in this feasibility study was deliberately kept as simple as possible. In particular the structure of the solid slab was constrained to be rigid. This meant that a number of interesting issues could not be addressed. One most important question is the competition with non-stoichiometric reconstructions observed for vacuum surfaces. One of the candidate structures for the rocksalt (111) surface is the so-called octopolar reconstruction suggested by Wolf\cite{Wolf1992} (see also Noguera\cite{Noguera2000,Goniakowski2008}). Dissolution of NaCl is a facile process interfering with reconstruction and the stability of polar surfaces is usually studied for more robust rocksalt crystals such as MgO\cite{Spagnoli2011} and NiO\cite{Freund1997}. 

However, under saturation the NaCl surface is in equilibrium with its aqueous solution and it has been proven to be feasible to stabilize a NaCl(111)/NaCl(aq) interface under appropriate thermodynamic conditions\cite{Radenovic2006}. The force field used in the present study (the Joung-Cheetham model of Ref.~\onlinecite{Joung2008}) is also suitable to carry out a simulation of a solid/solution equilibrium. After years of hard work there seems to be now a consensus in the computational literature regarding the solubility of this model (see the review of Ref.~\citenum{Nezbeda2016}). This is below the experimental value, but above the ~2 M concentration used in the present study for the solution phase. This ensures that there is no net tendency of the crystal to grow. These observations apply to non-polar surfaces (see however the results of Ref.~\citenum{Espinoza2016} for a spherical piece of crystal). It would therefore be of interest to repeat these calculations for a polar interface.  

Finally we return to the capacitance we computed in section \ref{sec:edlcap} for the polar surface electrolyte interface. The possibility of experimental realization of an unreconstructed polar surface electrolyte interface raises the question of the interpretation of this capacitance and how it might be observed. While the structure of the ``double layer'' should be a least in principle verifiable by experiment, the status of the corresponding capacitance is less clear. Capacitance is well defined for a nanoslab and should also be accessible to experiment. It would be the capacitance of the nanocapacitor formed by the solid slab with the electrolyte on either side acting as the two electrodes. But how to determine the capacitance of the interface of a non-conducting (insulator) semi-infinite crystal with a polar termination? The capacitance probably enters experimentally-measurable quantities only indirectly, such as in adsorption (complexation) energies of charged species. Maybe the double layer will also show electrokinetic signatures, such as a finite zeta-potential. This rather technical report is not the place to address these questions. This will have to be resolved in future investigations.

In conclusion we reiterate the observation by Noguera that the value of the compensating charge is a result of long range electrostatics and should remain the same even if the electronic structure is taken into account\cite{Noguera2000}. We therefore anticipate that this method can enable us to study the interaction of an electrolyte with more complex and realistic polar surfaces, possibly even treated by electronic structure calculation.

\begin{acknowledgments}
TS is supported by a departmental studentship (No.~RG84040) sponsored by the Engineering and Sciences Research Council (EPSRC) of the United Kingdom. The Research fellowship (No.~ZH 477/1-1) provided by the German Research Foundation (DFG) for CZ is gratefully acknowledged.
\end{acknowledgments}

\bibliography{DoubleLayers}

\end{document}